\documentstyle[preprint,aps,epsf,rotate]{revtex}
\begin{document}
\draft
%
\title{Calogero-Sutherland Model and Bulk-Boundary Correlations in
Conformal Field Theory}
\author{John Cardy}
\address{University of Oxford, Department of Physics -- Theoretical
         Physics, 1 Keble Road, Oxford OX1 3NP, U.K.\footnote{Address for
correspondence} \\
         and All Souls College, Oxford.}
%
%
\maketitle
\begin{abstract}
We show that, in any conformal field theory,
the weights of all bulk primary fields that couple to $N$
$\phi_{2,1}$ fields on the boundary are given by the spectrum
of an $N$-particle Calogero-Sutherland model.  The corresponding
correlation function is simply related to the $N$-particle wave function.
Applications are discussed to the minimal models and the non-unitary
O$(n)$ model.
\end{abstract}
%
%
The quantum
Calogero-Sutherland (C-S) model has proved to be ubiquitous in
theoretical physics. It has arisen in various ways in conformal field
theory (CFT) in the past\cite{cftcs}. 
In this note, we point out a very direct connection. This was
originally discovered\cite{JCdyson} in the course of developing a multiparticle
generalisation of Schramm-Loewner Evolution (SLE)\cite{SLE}, 
which, largely through the work of Lawler, Schramm and Werner (LSW)\cite{LSW},
has recently
enlarged our perspective on conformally invariant random processes.
However, the connection to the C-S model may be derived independently from
SLE, using the basic principles of CFT, and it 
will now be presented in this way.

The set-up is as follows: suppose we have a CFT in the interior of the unit
disc $|z|<1$, with a conformal boundary condition,
and consider in particular the correlation function 
\begin{equation}
\label{cf}
\langle\phi(e^{i\theta_1})\ldots\phi(e^{i\theta_N})\,\Phi(0)\rangle
=\langle\theta_1,\ldots,\theta_N|\Phi\rangle
\end{equation}
of $N$ boundary fields $\phi$ with a single primary bulk field
$\Phi$ at the origin. (Of course, any correlation function
in a simply connected region with $N$ boundary fields and a single
bulk field at an interior point may be
related to this by a conformal mapping.) In the second expression we have
written this correlation function in the operator formulation of CFT,
using radial quantisation: here $|\Phi\rangle$ is a highest weight state
of the holomorphic and antiholomorphic Virasoro algebras, and
$|\theta_1,\ldots,\theta_N\rangle$ is a state given by the action of
boundary operators $\phi(e^{i\theta_j})$ on the boundary state
corresponding to the given conformal boundary condition. These states
lie in an $N$-dimensional subspace of the full Hilbert space of the CFT. 

Let us now suppose that $\phi$ is a primary field which
is degenerate at level 2: it is a $\phi_{2,1}$ (or $\phi_{1,2}$) field
in the Kac classification. As shown many years ago by Belavin, Polyakov
and Zamolodchikov\cite{BPZ}, this implies that correlation functions such as 
(\ref{cf}) satisfy second-order differential equations. In this case we
shall show that these imply the C-S equation.

First fix some notation: parametrise the central charge by 
$c=1-6(4-\kappa)^2/4\kappa$, so that the boundary scaling dimension of
$\phi$ is $h_{2,1}=(6-\kappa)/2\kappa$, and the null vector
condition is $(L_{-2}-(\kappa/4)L_{-1}^2)|\phi_{2,1}\rangle=0$. 
Define the $N$-particle C-S hamiltonian with parameter $\beta$ by
\begin{equation}
H_N(\beta)\equiv
-\frac12\sum_{j=1}^N{\partial^2\over\partial\theta_j^2}
+{\beta(\beta-2)\over 16}\sum_{1\leq j<k\leq N}
{1\over\sin^2(\theta_j-\theta_k)/2}
\end{equation}
and the free fermion wave function 
\begin{equation}
\Psi_N(\theta_1,\ldots,\theta_N)=\prod_{1\leq j<k\leq N}
(e^{i\theta_j}-e^{i\theta_k})
\end{equation}
Then (subject to suitable boundary conditions, see later)
the ground state wave function of $H_N(\beta)$ is
$|\Psi_N|^{\beta/2}$ with energy $(\beta/2)^2E_N^{\rm ff}$
where $E_N^{\rm ff}=\frac1{24}N(N^2-1)$. 

Our main result is that if the correlation function (\ref{cf}) is
non-vanishing, then the highest weight (bulk scaling dimension) of 
$\Phi$ is given by
\begin{equation}
\label{eigen}
x_\Phi\equiv
h_\Phi+\overline h_\Phi=
{\kappa\over N}\Lambda_N(8/\kappa)-\frac4{N\kappa}E_N^{\rm ff}
+\frac{h_{2,1}}6+\frac c{12}
\end{equation}
where $\Lambda_N(\beta)$ is some eigenvalue of $H_N(\beta)$. Moreover, the 
correlation function (\ref{cf}) is (up to a normalisation) equal
to the corresponding eigenfunction, divided by  $|\Psi_N|^{2/\kappa}$.

Consider the infinitesimal conformal transformation 
$z\to z+\alpha(z)$, where $\alpha(z)=\sum_{j=1}^Nb_j\alpha_j(z)$, with 
\begin{equation}
\label{alphaj}
\alpha_j(z)=
-z{z+e^{i\theta_j}\over z-e^{i\theta_j}}
\end{equation}
where the $b_j$ are infinitesimal parameters, 
initially chosen to be arbitrary. 
Note that this preserves the unit circle with the points
$\{e^{i\theta_j}\}$ removed, but near the origin it acts as a pure
dilatation.
It may be implemented by inserting into the correlation function
(\ref{cf}) a factor $\int_CT(z)\alpha(z)dz/2\pi i
-\int_C\overline T(\bar z)\overline{\alpha(z)}d\bar z/2\pi i$,
where $T$ and $\overline T$ are the holomorphic and antiholomorphic
components of the stress tensor, and
$C$ is any contour in $|z|<1$ which encircles the origin once in a
counter-clockwise sense. 

The effect of this insertion may be evaluated in two ways: first by 
shrinking the contour towards the origin and observing that, as $z\to0$,
$\alpha(z)\sim z+O(z^2)$, and then using the operator product expansion
(OPE)
$T(z)\Phi(0)=h_\Phi\Phi(0)/z^2+O(z^{-1})$, together with a similar
antiholomorphic expression. This recovers the original correlation
function multiplied by $(h_\Phi+\overline h_\Phi)\sum_jb_j$.
It is important for this that $\Phi$ is primary, so that the
higher order terms in the expansion of $\alpha(z)$ do not contribute.

The other way is to distort $C$ so as to lie along the unit circle, with
small semicircles excluding the points $\{e^{i\theta_j}\}$. The
contributions from the parts of the contour on the circle vanish by
virtue of the conformal boundary condition\cite{JCsurf}. That from
the semicircle around $e^{i\theta_j}$ may be evaluated by writing
$z=e^{i\theta_j+i\zeta}$, where $\zeta$ is a local coordinate whose
imaginary part is zero along the boundary. Expanding $\alpha_j$ in
powers of $\zeta$, we find, after a little algebra,
$\alpha_j=b_j(2/\zeta-\zeta/6+O(\zeta^2))$.
Using the OPE with the stress tensor again, the effect of
the infinitesimal transformation $\alpha_j$ on $\phi(e^{i\theta_j})$
is to generate $-b_j(2L_{-2}-\frac16L_0)\phi(e^{i\theta_j})
=-b_j((\kappa/2)(\partial/\partial\theta_j)^2-\frac16h_{2,1})
\phi(e^{i\theta_j})$ (the minus sign is because $C$ wraps around 
$e^{i\theta_j}$ clockwise.) 
On the other hand, $\alpha_j$ is regular at the points $e^{i\theta_k}$
with $k\not=j$, so that
$\phi(e^{i\theta_k})\to (1+b_j\alpha'_j(e^{i\theta_k}))^{h_{2,1}}
\phi(e^{i\theta_k}+b_j\alpha_j(e^{i\theta_k}))$.

Putting together these contributions, the effect of $\alpha_j$ on the
boundary state is equivalent to
\begin{equation}
b_j\left[-{\kappa\over2}{\partial^2\over\partial\theta_j^2}
+{1\over6}h_{2,1}-
\sum_{k\not=j}\left(
\cot{\theta_k-\theta_j\over2}{\partial\over\partial\theta_k}
+i\cot{\theta_k-\theta_j\over2}h_{2,1}
-{1\over2\sin^2(\theta_k-\theta_j)/2}h_{2,1}\right)\right]
\end{equation}

Now sum over $j$: the penultimate terms in the above sum to something 
proportional to $\sum_j\sum_{k\not=j}(b_j+b_k)\cot(\theta_k-\theta_j)/2$,
which vanishes if we now take $b_j=b_k$ for each pair $(j,k)$. 
The generator of the transformation, acting in the subspace of boundary
states, is therefore
\begin{equation}
\label{G}
G\equiv -\frac\kappa2\sum_j{\partial^2\over\partial\theta_j^2}
-\sum_j\sum_{k\not=j}
\cot{\theta_k-\theta_j\over2}{\partial\over\partial\theta_k}
+\left(\sum_j\sum_{k\not=j}{1\over2\sin^2(\theta_k-\theta_j)/2}
+\frac N6\right)h_{2,1}
\end{equation}
The first two terms can be recognised as a similarity transform 
of $H_N(4/\kappa)$, up to a constant: 
\begin{equation}
|\Psi_N|^{2/\kappa}G|\Psi_N|^{-2/\kappa}
=\kappa\left[H_N(4/\kappa)-(2/\kappa)^2E_N^{\rm ff}\right]
+\left(\sum_j\sum_{k\not=j}{1\over2\sin^2(\theta_k-\theta_j)/2}
+\frac N6\right)h_{2,1}
\end{equation}
The penultimate term then combines with the potential term in
$H_N(4/\kappa)$ to give a Calogero-Sutherland hamiltonian at a
\em shifted \em value of $\beta$, equal to $8/\kappa$:
\begin{equation}
|\Psi_N|^{2/\kappa}G|\Psi_N|^{-2/\kappa}
=\kappa H_N(8/\kappa)-(4/\kappa)E_N^{\rm ff}+\frac N6h_{2,1}
\end{equation}
The eigenvalues and eigenvectors of $G$ and $H_N(8/\kappa)$ are thus
simply related, and since the left eigenvalues of $G$ are $N(h_\Phi
+\overline h_\Phi)$, we get (\ref{eigen}), except for the last term.
This arises because the same insertion must be made in the partition
function, which transforms non-trivially owing to the presence of a
non-zero trace $\langle\Theta\rangle$ of the stress tensor at the curved
boundary\cite{CP}. However, a clearer derivation of this term 
may be found by considering the conformally equivalent geometry of a
semi-infinite cylinder parametrised by the complex variable $\ln z$. The
boundary is now no longer curved, but the dilatation operator becomes
the generator of translations along the cylinder, which is\cite{JCopc} 
$L_0+\overline L_0-c/12$.

We now discuss some examples and applications of the general result.
The C-S hamiltonian $H_N$ commutes with the total momentum $P\equiv
-i\sum_j(\partial/\partial\theta_j)$, whose eigenvalues correspond to
the spin $h_\Phi-\overline h_\Phi$ of the bulk primary
field. The eigenvalue equation admits two possible boundary
conditions on the wave function $\psi$
as a given pair $(j,k)$ of particles approach each other:
$\psi\propto|\theta_j-\theta_k|^\gamma$, with $\gamma=\beta/2$ (`fermionic')
or $1-\beta/2$ (`bosonic'). 
These correspond to the values allowed by the BPZ fusion
rules\cite{BPZ}: the correlation function behaves as
$|\theta_j-\theta_k|^{\gamma-2/\kappa}$, which is consistent with the OPE
\begin{equation}
\label{ope}
\phi_{2,1}(\theta_j)\cdot\phi_{2,1}(\theta_k)
\sim |\theta_j-\theta_k|^{-2h_{2,1}}\,{\bf 1}
+|\theta_j-\theta_k|^{h_{3,1}-2h_{2,1}}\,\phi_{3,1}
\end{equation}
with $h_{2,1}=(6-\kappa)/2\kappa$ and 
$h_{3,1}=(8-\kappa)/\kappa$. 

The eigenvalues of $H_N(\beta)$ in the fermionic case are 
well-known\cite{Suth}: they have the form  $\Lambda=\frac12\sum_{j=1}^Nk_j^2$,
where the allowed values of the quasiparticle momenta $k_j$ satisfy
$\sum_jk_j=P$ and
$k_{j+1}-k_j=\frac12\beta+p_j$, with $p_j$ a non-negative integer.
The ground state in this sector has all the $p_j=0$, and 
eigenvalue $(\beta/2)^2E_N^{\rm
ff}$, which, after a little algebra, leads to a weight
\begin{equation}
x_N^{\rm f}={N^2\over2\kappa}-{(4-\kappa)^2\over 8\kappa}
\end{equation}
of the corresponding bulk field. 

\subsubsection*{The O$(n)$ model.}
The most immediate example of such a bulk field is in the
non-unitary, non-minimal
CFT that is supposed to represent the scaling limit
of the O$(n)$ model\cite{DF} with $n\in[-2,2]$. This may be realised
as a gas of non-intersecting closed loops 
and open curves\cite{Nien}, in which open curves
ending on the boundary are known to be described by $\phi_{2,1}$
fields\cite{JCsurf}. The above result for $x_N^{\rm f}$ then
agrees with the known value\cite{DupSal}
for the bulk $N$-leg field. That is, the correlation function
(\ref{cf}) is proportional to the probability that $N$ non-intersecting
curves connect the origin to the points $e^{i\theta_j}$ on the
boundary. 

For $N=1$ and $P\not=0$ we find a weight $x_\Phi=x_1^{\rm f}+(\kappa/2)P^2$
with $P$ an integer. The physical interpretation of these new primary
fields is in terms of winding number states: each curve linking the origin
and the boundary with winding number 
$\chi$ is weighted by $e^{iP\chi}$. 

For $N=2$, in addition to the winding states with $P\not=0$, there are 
new spinless primary fields corresponding to $k_2=-k_1=\beta/4+p$,
with weights $x_\Phi=x_2^{\rm f}+\frac12\kappa p^2+2p$, 
where $p$ is a positive
integer: these correspond in the O$(n)$ model to excited modes of the
pair of curves, confined by their mutual repulsion.
In the case of a finite cylinder, or annulus, their weights give the
exponents of correction terms in the correlation function.
There are further $N=2$ primary fields corresponding to
purely bosonic boundary conditions, with
$\gamma=1-\beta/2$. These
correspond to $\Lambda_2=(1-\beta/2)^2E_2^{\rm ff}$,
which gives $x_2^{\rm b}=0$ for all $\kappa$. This is
consistent with the OPE (\ref{ope}): the two boundary fields are
fusing to the identity on the boundary, which then couples to the
identity field in the bulk. However, once again there are excited
states in this sector, which correspond to possible new bulk primary
fields with weights $x_\Phi=\frac12\kappa p(p+1)-2p$. These two types of
boundary condition may be understood within the O$(n)$ model as follows:
each boundary field $\phi_{2,1}$ and its attached curve carry an 
O$(n)$ vector index. If the two
labels are different, the curves cannot join, and the fusion is
into a $\phi_{3,1}$ field transforming according to a tensor
representation of O$(n)$. However, if the labels are the same, the
curves can join up before reaching the origin. The fusion in this
case is into the identity field, and the fact that the leading
coupling is now to a bulk primary field with $x_\Phi=0$ means that
the probability of the curves joining, and not therefore passing
thorugh the origin,
is unity. The weights corresponding to the excited states, with
$p\geq1$, then give the
exponents of correction terms to this for an annulus. 
$p=1$ corresponds to the bulk energy density field of the O$(n)$ model.

However, still for $N=2$, there are other possible `mixed' boundary
conditions which are fermionic as $\theta_2-\theta_1\to0+$, and bosonic
as $\theta_2-\theta_1\to2\pi-$. The ground state in this sector has
energy $\Lambda_2=\frac1{16}$, which gives 
$x_\Phi=(3\kappa-8)(8-\kappa)/32\kappa$. This is the exponent determining
the relative probability that a curve, whose ends are 
attached at nearby points on the
boundary, should enclose the origin or not. For $\kappa=6$, this is the
`one-arm' exponent of percolation (related to the probability that the
origin lies in a cluster which touches the boundary), as computed by
LSW\cite{LSW1arm}.
Once again, there are excited states in this sector whose
energies give corrections to scaling.

\subsubsection*{Minimal models.}
These correspond to rational $\kappa=4k/k'$, and the allowed values of the
weights of scalar bulk primary fields
are given by the Kac formula
$x_{r,s}=((4r-\kappa s)^2-(4-\kappa)^2)/8\kappa$ with $1\leq r\leq k-1$
and $1\leq s\leq k'-1$. For these to agree with (\ref{eigen}) imposes a
severe constraint. If it cannot be satisfied, it implies that the
correlation function (\ref{cf}) must vanish. For example, the weight of
the $1$-leg field corresponds formally to $x_{1/2,0}$, but for this to
appear in the table of allowed values it is necessary that $k$ is odd,
with $r=(k-1)/2$, and $k'$ even, with $s=k'/2$. This is of course
consistent with the fusion rules of boundary conformal field theory. 
For $N=2$, with fermionic boundary conditions,
no solution for $(r,s)$ in the allowed range is possible,
indicating that, in a minimal model, two $\phi_{2,1}$ fields on the
boundary can couple to the bulk only through fusion into the identity.
In that case, coupling is allowed, as long as $p\leq[k/2]-1$,
because the corresponding allowed weight is $x_\Phi=x_{1,2p+1}$.

\subsubsection*{Comparison with multiple SLEs.}
Although the above CFT arguments are self-contained, it is instructive
to compare them with those of Ref.~\cite{JCdyson}. 
There, a multi-particle generalisation of SLE was proposed in which the
$N$ curves connecting the boundary with the origin are `grown'
dynamically, starting from the boundary at time $t=0$ and reaching the
origin as $t\to\infty$. This process is described in terms of 
the evolution of the conformal
mapping $g_t(z)$ which sends the simply connected region not yet excluded
by the curves into the whole unit disc. This turns out to satisfy
$dg_t=\sum_jb_j\alpha_j(g_t)$, with $\alpha_j$ 
having the same
form as in (\ref{alphaj}). In this picture, however, the $\theta_j$ become
functions of $t$, evolving, if we take $b_j=dt$ for all $j$,
according to Dyson's brownian motion\cite{dyson}:
\begin{equation}
\label{dysonbm}
d\theta_j=\sum_{k\not=j}\cot((\theta_j-\theta_k)/2)\,dt+dB_j(t)
\end{equation}
The terms on the right hand side correspond to a mutual repulsion and a
stochastic noise. For $N=1$ it is known\cite{SLE} that this process
(radial SLE) gives the correct measure on the continuum limit of a
single curve. The corresponding Fokker-Planck equation for the joint
probability distribution $P(\{\theta_j\};t)$ has the form 
$dP/dt={\cal L}P$, and it was argued in Ref.~\cite{JCdyson} that the
asymptotic equilibrium solution of this equation, satisfying ${\cal L}P_{\rm
eq}=0$, should
give the distribution of the points on the boundary in an equilibrium
2$d$ critical system such as the O$(n)$ model. This gives the result
$|\Psi_N|^{4/\kappa}$, in contradiction with the CFT prediction
$|\Psi_N|^{2/\kappa}$.

This discrepancy may be traced to the form of 
\begin{equation}
{\cal L}^{\dag}={\kappa\over2}\sum_j{\partial^2\over\partial\theta_j^2}
+\sum_j\sum_{k\not=j}\cot{\theta_k-\theta_j\over2}{\partial\over\partial
\theta_k}
\end{equation}
The first term comes from averaging over the white noise $dB_j$, and the
second from the repulsion. Comparing this with the expression (\ref{G})
for the conformal generator $G$, we see that the last term, proportional
to $h_{2,1}$, is absent. This arises, in the CFT calculation,
from the transformation of
$\phi(e^{i\theta_k})$ under the local scale
transformation induced by the conformal mapping $\alpha_j$, but it is
missing in the multiple SLE approach. Indeed, there is no obvious way of
modifying (\ref{dysonbm}) so as to incorporate such a term.

However, the argument in Ref.~\cite{JCdyson} that the joint distribution of the
boundary points 
$\{\theta_j\}$ in a critical system should be given by the equilibrium
distribution
of the process (\ref{dysonbm}) was also based on the assumption that the
measure on the $N$ curves was strictly conformally invariant under
$g_t$. For the
case of a single curve with each end on the boundary of a simple
connected region, this is known not to be the case in general -- rather, it is
conformally covariant.\cite{LSW}
The covariance factor is just the product of the
local scale transformations at each end, raised to the power $h_{2,1}$.
When the arguments of Ref.~\cite{JCdyson} are modified to take into
account these factors, the result agrees precisely with that of
the CFT argument given earlier.
The fact that SLE$_\kappa$ is now associated with $H_N(\beta)$ with
$\beta=8/\kappa$ is much more satisfactory. For example, it suggests
that the duality of SLE\cite{SLEduality}
under $\kappa\to16/\kappa$ is related to the
known duality of the C-S hamiltonian\cite{CSduality} under $\beta\to4/\beta$. 

\vspace{1cm}
To summarise, we have shown that the quantum Calogero-Sutherland model
arises in a very simple way in bulk-boundary conformal field theory. 
The full spectrum is realised in the non-unitary CFT of the
O$(n)$ model, and it predicts the scaling dimensions of new primary
fields in that theory. In minimal models, it places severe
restrictions on which bulk fields can couple to the boundary. 
In general, because of the Galilean invariance of the C-S hamiltonian,
any scalar primary bulk operator is associated with a tower of other
primaries of spin $s$, with the differences in conformal weights proportional
to $s^2$. Thus, although the spectrum of Virasoro descendants in a CFT
is relativistic, the spectrum of these primaries has a non-relativistic form.

Although we have considered only the case of $\phi_{2,1}$ fields on
the boundary, our arguments can be generalised to include other
boundary fields corresponding to degenerate Virasoro representations.
These will lead to higher-order differential operators. Similar
generalisations to WZWN models are also possible.

Recently Bauer and Bernard have
extended their analysis of CFT as a probe of SLE\cite{BB} to the
radial case\cite{BBrad}. Some of their results overlap with ours. 

The author acknowledges both helpful and critical conversations with 
D.~Bernard, G.~Lawler and W.~Werner. This work was supported in part
by the EPSRC under Grant GR/J78327.

%
%
%
\end{document}